\documentclass[aps,pre,reprint,amsmath,amssymb,superscriptaddress, footinbib]{revtex4-2}
\usepackage{graphicx}
\usepackage{color}
\usepackage{amsfonts}
\usepackage{physics}
\usepackage{braket}
\usepackage{multibib}
\usepackage[english]{babel}
\usepackage{bm}
\usepackage[colorlinks=true, allcolors=blue]{hyperref}

\begin{document}
\title{A link between anomalous viscous loss and boson peak in soft jammed solids}
\author{Yusuke Hara}
\email{yusukephy@gmail.com}
\affiliation{Graduate School of Arts and Science, The University of Tokyo, Komaba, Tokyo 153-8902, Japan}
\author{Ryosuke Matsuoka}
\affiliation{Department of Physics, Kyushu University, 744 Motooka, Nishi-ku, Fukuoka 819-0395, Japan}
\author{Hiroyuki Ebata}
\affiliation{Department of Physics, Kyushu University, 744 Motooka, Nishi-ku, Fukuoka 819-0395, Japan}
\author{Daisuke Mizuno}
\email{mizuno@phys.kyushu-u.ac.jp}
\affiliation{Department of Physics, Kyushu University, 744 Motooka, Nishi-ku, Fukuoka 819-0395, Japan}
\author{Atsushi Ikeda}
\email{atsushi.ikeda@phys.c.u-tokyo.ac.jp}
\affiliation{Graduate School of Arts and Science, The University of Tokyo, Komaba, Tokyo 153-8902, Japan}
\affiliation{Research Center for Complex Systems Biology, Universal Biology Institute, The University of Tokyo, Komaba, Tokyo 153-8902, Japan}

\date{\today}

\begin{abstract}
Soft jammed solids exhibit intriguing mechanical properties, while their linear response is elusive. In particular, foams and emulsions generally reveal anomalous viscous loss with the loss and storage modulus following $G^{\prime \prime} \propto \sqrt{\omega}$ and $G^{\prime} \propto \omega^0$.
In this study, we offer a comprehensive microscopic understanding of this behavior.
Using microrheology experiment, we measured $G^* = G^{\prime} + i G^{\prime \prime}$ of concentrated emulsions in a wide range of frequencies.
In theory, we applied a linear response formalism for microrheology to a soft sphere model that undergoes the jamming transition.
We find that the theory quantitatively explains the experiments without the need for parameter adjustments.
Our analysis reveals that the anomalous viscous loss results from the boson peak, which is a universal vibrational property of amorphous solids and reflects the marginal stability in soft jammed solids.
We discuss that the anomalous viscous loss is universal in systems with various interparticle interactions as it stems from the universal boson peak, and it even survives below the jamming density where thermal fluctuation is pronounced and the dynamics becomes inherently nonlinear.
\end{abstract}

\maketitle



Soft jammed solids, composed of densely packed mesoscopic or macroscopic particles, are ubiquitous in nature and modern life, ranging from foods, pastes, cosmetics, and soils to mudflows. 
Mechanical properties of these materials vary highly sensitively depending on factors such as density, microstructure, deformation history, and so on~\cite{Bonn2017,Morris2020,Behringer_2019}. 
To understand the mechanism underlying these behaviors, researchers must first examine the linear viscoelasticity, a response to infinitesimal deformation~\cite{larson1999structure}. 
This is essential because it directly probes the microscopic and mesoscopic dynamics, reflecting the internal structure of the material.
Unlike many other materials, however, understanding the linear viscoelasticity, $G^{\prime}$ (storage modulus) and $G^{\prime \prime}$ (loss modulus), has already been challenging for soft jammed solids.
Experimental studies have shown that the complex modulus $G^* = G^{\prime} + i G^{\prime \prime}$ of soft jammed materials exhibits anomalous power-law dependence on frequency~\cite{mason1995, liu1996, cohen1998, hebraud2000role, gopal2003, bandyopadhyay2006slow, marze2008aqueous,besson2008,krishan2010,kropka2010,gupta2012advanced,basu2014rheology, hanotin2015,nishizawa2017(a), nishizawa2017(b), conley2019relationship}.
In particular, $G^{\prime} \propto \omega^0$ and $G^{\prime \prime} \propto \sqrt{\omega}$ have been observed widely in emulsions, foams, microgels, and cytoplasm~\cite{mason1995, liu1996, cohen1998, hebraud2000role, gopal2003, besson2008, krishan2010, kropka2010,gupta2012advanced, basu2014rheology, nishizawa2017(a), nishizawa2017(b), conley2019relationship}. 
This phenomenon, referred to as the anomalous viscous loss~\cite{liu1996}, indicates that the dissipation in soft jammed solids cannot be explained using common Hookean or Kelvin-Voigt solids, even in apparently elastic regime.

The theory of the jamming transition has been developed to provide a microscopic basis to understand the mechanical properties of soft jammed solids~\cite{Behringer_2019,van2009,Liu2020}.
In this framework, the system is described by a soft sphere model, an athermal assembly of short-range, purely repulsive particles.
With increasing the density, the system acquires a finite rigidity at the density $\phi_J$, which is the jamming transition~\cite{ohern2003}. 
Simulations and theories have established that soft sphere models exhibit critical power-law scalings of geometrical, vibrational, and mechanical properties near $\phi_J$~\cite{ohern2003, Wyart2005, van2009, parisi2020theory}.
For static properties, such as the contact number per particle~\cite{Katgert2010} and the static shear modulus~\cite{lacasse1996}, these predictions show consistency with the experimental results at the density $\phi > \phi_J$ where the athermal assumption is appropriate.
However, such an agreement has yet to be achieved for the foundational mechanical property, i.e., $G^*$.
Because the linear response in an athermal system is governed by the Hessian matrix of the potential function, the complex modulus $G^*$ can be deduced by using the vibrational eigenmodes and eigenfrequencies of the system~\cite{lemaitre2006,tighe2011}. 
For the soft sphere model, this approach predicted the critical power-law behavior $G^{\prime} = G^{\prime \prime} \propto \sqrt{\omega}$ in the high-frequency regime and the elastic plateau $G^{\prime} \propto \omega^0$, $G^{\prime \prime} \propto\omega$ in the low-frequency regime~\cite{tighe2011}. 
These predictions, however, do not coincide with the experimental observations mentioned earlier, even at the level of the scaling laws. 
As a result, understanding of the anomalous viscous loss remains phenomenological~\cite{liu1996}.

This work aims at a comprehensive microscopic understanding of $G^*$ in soft jammed solids.
Experimentally, we perform microrheology measurements for concentrated emulsions to obtain a broadband spectrum of $G^*$ beyond the range accessible to macrorheology. 
Theoretically, we apply a linear response formalism for microrheology~\cite{hara2023} to a soft sphere model. 
We find that the theory quantitatively reproduces the experimental $G^*$ at $\phi > \phi_J$ without adjustable parameters, pinpointing the origin of the anomalous viscous loss. 
This loss directly results from the boson peak, a universal vibrational property of amorphous solids overlooked in the previous analysis~\cite{tighe2011}.
Moreover, the theoretical scaling function explains the experimental $G^*$ even at $\phi < \phi_J$ where thermal fluctuation is pronounced and the dynamics becomes inherently nonlinear. 
Based on these results and additional simulations, we discuss the universality of the anomalous viscous loss in soft jammed solids.

\begin{figure}[t]
\centering 
\includegraphics[width=0.95\linewidth]{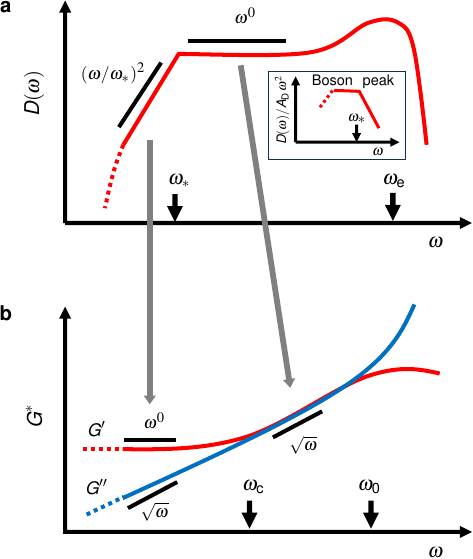}
\caption{
{\bf The link between the vibration and the linear viscoelasticity revealed in this work.} 
{\bf a.} A schematic picture of the vibrational density of states of the soft sphere model, which has been established previously. 
The plateau $D(\omega) \propto \omega^0$ emerges at the intermediate-frequency $\omega_* < \omega < \omega_e$, and the non-Debye scaling law $D(\omega) \propto (\omega/\omega_*)^2$ at the low-frequency $\omega < \omega_*$.  
The inset shows $D(\omega)/A_{\rm D} \omega^2$. 
The non-Debye scaling regime constitutes the boson peak. 
{\bf b.} A schematic picture of the complex modulus of the soft sphere model, which was obtained in this work. 
The critical power-law behavior $G^{\prime} = G^{\prime \prime} \propto \sqrt{\omega}$ emerges at $\omega_c < \omega < \omega_0$ and the anomalous viscous loss $G^{\prime} \propto \omega^0, G^{\prime \prime} \propto \sqrt{\omega}$ at $\omega < \omega_c$. 
The arrows between the panels indicate the causal relationships. 
The characteristic frequencies are related as $\omega_0=\omega_e^2$ and $\omega_c=\omega_*^2$, where the square comes from a time scale relation between the inertial dynamics in vibration and the overdamped dynamics in viscoelasticity. 
}
\label{fgr:concept}
\end{figure}

\section*{Background on vibrational properties of the soft sphere model}

``Hard'' structural glasses, such as inorganic and molecular glasses, universally share an excess of low-frequency vibrations~\cite{phillips1981amorphous}.  
Namely, the vibrational density of states (vDOS) $D(\omega)$, the number of vibrational modes at a given frequency $\omega$, becomes significantly larger than $A_{\rm D} \omega^2$, the prediction by the classical Debye theory. 
As a result, $D(\omega)/A_{\rm D} \omega^2$ shows a pronounced peak in the low-frequency regime, called the boson peak. 
The boson peak is crucial as it dramatically modifies the solid state properties of glasses~\cite{phillips1981amorphous}. 
While various optical, thermal, and scattering measurements have established the presence of the boson peak, its microscopic origin has long been debated~\cite{phillips1981amorphous,Schirmacher2007,Klinger2010}.

Recent studies showed that the boson peak is also present in soft jammed solids, and it can be explained well by the marginal stability of the systems.
Although the motion of particles in many soft jammed solids is overdamped, it is theoretically beneficial to consider a hypothetical inertial motion of particles and study their vibrations~\cite{Lin2016}.
The so-obtained ``vDOS'' of the soft sphere model is schematically shown in Fig.~\ref{fgr:concept}a.
At higher frequency $\omega > \omega_*$, the vDOS becomes flat $D(\omega) \propto \omega^0$, which we call the plateau part.
As the system approaches the jamming point, the characteristic frequency $\omega_*$ vanishes, and the plateau develops~\cite{silbert2005, silbert2009}.

At lower frequency $\omega < \omega_*$, the vDOS obeys $D(\omega) \propto (\omega/\omega_*)^2$, which is called the non-Debye scaling law~\cite{degiuli2014a,charbonneau2016,Lin2016,mizuno2017}.
Though similar, this law differs widely from the Debye law $D(\omega) = A_{\rm D} \omega^2$.
(i) The vibrational modes in this regime are not plane wave~\cite{mizuno2017}.
(ii) Unlike the Debye law, the exponent 2 is independent of the spatial dimension~\cite{charbonneau2016,shimada2020low}.
(iii) The coefficient $\omega_*^{-2}$ is much larger than $A_{\rm D}$, and $\omega_*^{-2}/A_{\rm D}$ even diverges at the jamming point~\cite{degiuli2014a,mizuno2017}.
As a result, the non-Debye scaling regime constitutes a huge boson peak of the soft sphere model (Fig.~\ref{fgr:concept}a, inset).
Theoretically, this law is a consequence of the marginal stability of the system~\cite{degiuli2014a,franz2015}: the system is on the verge of mechanical instability against the compressive forces between particles~\cite{degiuli2014a}.
In this work, we show that this non-Debye scaling regime is responsible for the anomalous viscous loss in the linear viscoelastitiy.

Note that at much lower frequency $\omega \ll \omega_*$, the vDOS deviates from the non-Debye scaling law, and the vibrational modes become a mixture of the plane wave and the quasi localized modes~\cite{Lerner2016,mizuno2017}.

\begin{figure}[t]
\includegraphics[width=0.95\linewidth]{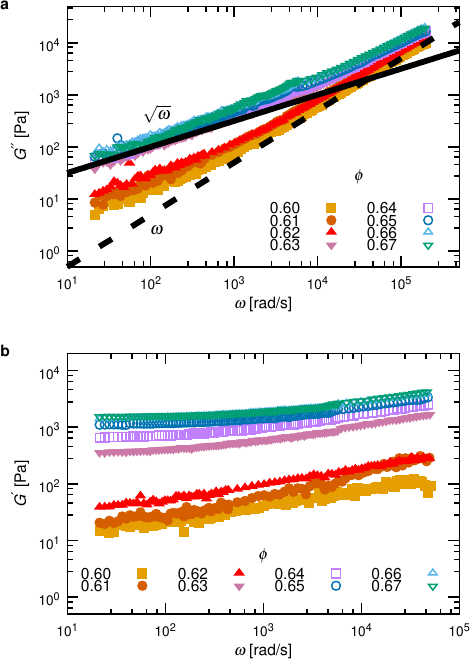}
\caption{
{\bf The complex modulus of the concentrated emulsions obtained by the microrheology experiments.} 
{\bf a.} The loss modulus for the packing fraction $\phi = 0.60-0.67$. 
The dashed line $G^{\prime \prime} \propto \omega$ describes the high-frequency behavior, while the solid line $G^{\prime \prime} \propto \sqrt{\omega}$ describes the low-frequency behavior. 
{\bf b.} The storage modulus for the same densities, which converges to the density-dependent finite values in the low-frequency regime. 
}
\label{fgr:exp-modulus}
\end{figure}

\section*{Microrheology experiment}

A broadband spectrum of the complex modulus of soft jammed solids is necessary for a quantitative comparison between theory and experiment.
We achieved this by performing microrheology experiments on concentrated oil-in-water emulsions prepared by slightly modifying the procedure detailed in a previous study~\cite{senbil2019}.  
In particular, we performed passive microrheology, which involves tracking the spontaneous thermal motion of a probe particle with the radius $a$. 
To measure the complex modulus $G^*(\omega)$ over a wide frequency range, we employed a high-bandwidth laser particle tracking technique~\cite{schnurr1997}. 
We defined the complex response function for a tracer particle as $\alpha^*(\omega) = \tilde{u}(\omega)/\tilde{F}(\omega)$, which represents the ratio of the applied sinusoidal force $F(t) = \tilde{F}(\omega) e^{i\omega t}$ to the resulting displacement response $u(t) = \tilde{u}(\omega) e^{i\omega t}$. 
In the passive microrheology, we derive the response function from the observed thermal fluctuation of the probe particle by employing the fluctuation-dissipation theorem. 
By extending the Stokes' formula to a frequency domain, i.e., $G^* = G^{\prime} + i G^{\prime \prime} = 1/(6 \pi a \alpha^*)$, we obtained the complex shear modulus of the medium surrounding the probe particle. 
Further details are provided in Methods. 
    
Figure~\ref{fgr:exp-modulus} shows the measured loss and storage modulus at various packing fractions $\phi=0.60-0.67$. 
While their absolute magnitudes increase with density, their frequency dependence is similar among different densities. 
The loss modulus $G^{\prime \prime}$ is proportional to $\omega$ in the high-frequency regime and $\sqrt{\omega}$ in the low-frequency regime. 
The storage modulus $G^{\prime}$ shows power-law frequency dependence in the high-frequency and converges into the density-dependent finite value in the low-frequency.
Therefore in the low-frequency regime, the anomalous viscous loss $G^{\prime} \propto \omega^0, G^{\prime \prime} \propto \sqrt{\omega}$ emerges at all the densities studied.
We emphasize that the microrheology technique lets us simultaneously observe the high- and low-frequency regimes.
The following sections build a microscopic and quantitative understanding of these experimental results.

\begin{figure*}[t]
    \includegraphics[width=\linewidth]{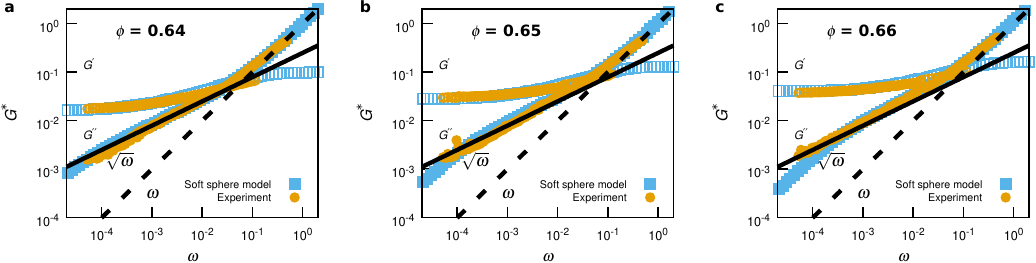}
    \caption{
{\bf Direct comparison between the theory and the experiment.}  
{\bf a.b.c.} The experimentally measured complex modulus of the concentrated emulsion (circles) and the numerically computed complex modulus of the soft sphere model (squares). 
The solid and dashed lines represent $G^{\prime \prime} \propto \sqrt{\omega}$ and $G^{\prime \prime} \propto \omega$. 
The results for three different packing fractions are displayed: $\phi = 0.64$ ({\bf a}), $0.65$ ({\bf b}), and $0.66$ ({\bf c}). 
The comparison is virtually free from adjustable fitting parameters. 
}
\label{fgr:exp-model}
\end{figure*}

\section*{Numerical analysis of the soft sphere model}

We introduce a soft sphere model to study the complex modulus of emulsions numerically and theoretically. 
Here, we describe our model and numerical results focusing on $\phi>\phi_J$.

Our soft sphere model is composed of $N$ three-dimensional particles interacting through the pairwise potential
\begin{equation}
   v(r_{ij}) = \frac{\sigma D^2 \beta}{2} \left(\left(\frac{D}{r_{ij}} \right)^{3} - 1\right)^{\alpha} \Theta\left(\frac{D}{r_{ij}} - 1\right),
   \label{eq:potential}
\end{equation}
where $r_{ij}$ express the interparticle distance. 
We set $\alpha=2.23$ and $\beta=0.26$ so that the potential accurately describes the interaction between droplets near the jamming transition~\cite{lacasse1996}.
The diameter and the interfacial tension of the droplets are $D=4.8 \times 10^{-7}$m and $\sigma=8.8 \times 10^{-3}$N/m, whose values are taken from the experiments (See Methods).

We prepare a mechanically stable configuration of the particles in a simulation box and apply an oscillatory force to one of the particles (a probe particle).
The equation of motion is 
\begin{equation}
  C_{0} \sum_{j \in \partial_i}\left(\dv{\vec{r}_i}{t} - \dv{\vec{r}_j}{t}\right) = - \pdv{U}{\vec{r}_i} + \vec{F}_p \delta_{ip}. 
\label{eq:eom}
\end{equation}
In the right-hand side, $U = \sum_{i > j} v(r_{ij})$ is the total potential and $\vec{F}_p$ is an infinitesimally small sinusoidal force applied to the probe particle. 
We omitted the thermal agitation as the contact force is more significant at $\phi > \phi_J$.
The left-hand side models the contact dissipation between particles, where $\partial_i$ indicates a set of the neighbors of particle $i$, and $C_0$ is the dissipation coefficient.
We solved this equation of motion at the level of linear response and tracked the motion of the probe particle to obtain the response function $\alpha^*(\omega)$: its explicit formula is presented in Eq.~(\ref{eq:response-function-LR}). 
We then employed the generalized Stokes' formula to convert $\alpha^*(\omega)$ into the complex modulus $G^*(\omega)$. 
Further details are provided in Methods.

Two considerations are required when we compare numerical and experimental results.
First, we should compare the results not at the same packing fraction $\phi$ but at the same distance to the jamming density $\phi-\phi_J$.
This is because the dependence of various physical quantities on $\phi-\phi_J$ are robust~\cite{ohern2003} while $\phi_J$ sensitively depends on the polydispersity~\cite{ohern2003} and the preparation protocol~\cite{Ozawa2017}, which are hard to control. 
Indeed, the estimated jamming density is $\phi_J = 0.628$ for our experiments and $\phi_J = 0.638$ for our simulations. 
Second, we need to determine the contact dissipation coefficient $C_0$, which may also depend on the microscopic details of the droplets. 
Here, we measured the high frequency viscosity $\eta_{0} = \lim_{\omega \to \infty} G^{\prime \prime}(\omega)/\omega$ in the experiments, and set $C_0 = 2.0 D \eta_0$ so that the model reproduces it.
The high-frequency resolution of the microrheology technique enabled this determination protocol.
After these considerations, there is no remaining adjustable parameter for the comparison.

Figure~\ref{fgr:exp-model} shows the direct comparison at $\phi>\phi_J$.
Hereafter, we nondimensionalize all the quantities using $R=D/2$, $\sigma R^2$, and $R\eta_0/\sigma$ as the length, energy, and time units, respectively.
In all the densities and all the frequency regimes, the agreement between the numerical and experimental results is remarkable.
This agreement establishes that the soft sphere model quantitatively reproduces the linear viscoelasticity of the concentrated emulsions, also pointing that the anomalous viscous loss $G^{\prime} \propto \omega^{0}, G^{\prime \prime} \propto \sqrt{\omega}$ can be explained within the soft sphere model.

\begin{figure*}[t]
    \includegraphics[width=\linewidth]{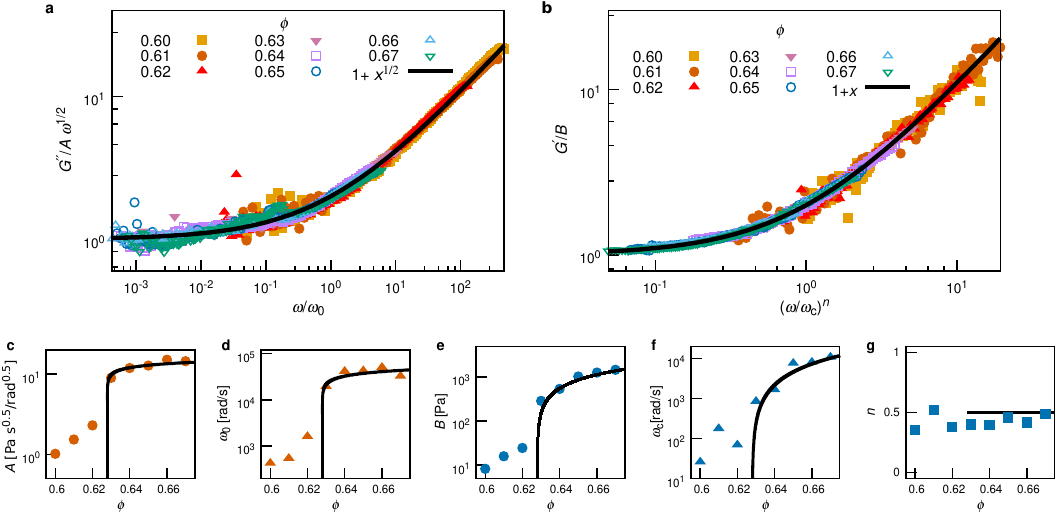}
    \caption{
{\bf Scaling analysis of the experimentally measured complex modulus.} 
{\bf a.b.} The loss modulus $G^{\prime \prime}$ ({\bf a}) and the storage modulus $G^{\prime}$ ({\bf b}) for $\phi=0.60-0.67$ are plotted according to the scaling functions Eq.~(\ref{eq:fitting}). 
The density-dependent parameters $A, B, \omega_0, \omega_c, n$ were determined so that each data set collapsed onto the master curve. 
{\bf c.d.e.f.g.} The determined parameters $A, B, \omega_0, \omega_c, n$ are plotted against the packing fraction $\phi$. 
The solid lines represent the theoretical predictions for $\phi > \phi_J$:  $A \propto \left(\phi-\phi_J\right)^{\left(\alpha-2\right)/2}$ ({\bf c}), $B \propto \left(\phi-\phi_J\right)^{\alpha-3/2}$ ({\bf d}), $\omega_0 \propto \left(\phi-\phi_J\right)^{\alpha-2}$ ({\bf e}), $\omega_c \propto \left(\phi-\phi_J\right)^{\alpha-1}$ ({\bf f}), and $n=1/2$ ({\bf g}), where $\phi_J = 0.628$ is the jamming density of the experimental system. 
}
    \label{fgr:scaling-modulus}
\end{figure*}

\section*{Theoretical analysis of the soft sphere model}

We now build a microscopic understanding of the complex modulus of the emulsions. 
Since we focus on the scaling behaviors, we omit unimportant $O(1)$ quantities in this section. 

As shown in Methods, the linear response formalism enables us to express the complex modulus $G^*(\omega)$ in terms of the vibrational properties of the system:
\begin{equation}
\frac{1}{G^*(\omega)} = \int d\omega^{\prime} \frac{D(\omega^{\prime})}{\left(\omega^{\prime}\right)^2 + i \omega}, 
\label{eq:modulus-dos}
\end{equation}
where $D(\omega)$ is the vDOS of the soft sphere model.
As discussed earlier, $D(\omega)$ is characterized by the plateau and the non-Debye scaling law (Fig.~\ref{fgr:concept}a, see also Fig.~\ref{fgr:scaling-dos}):
\begin{equation}
D(\omega) = 
\begin{cases}
1/\omega_e
& \left(\omega_{*} < \omega < \omega_e \right) \\
\omega^2 / \left( \omega_e \omega_*^2 \right) 
& \left(\omega < \omega_*\right). 
\label{eq:vDOS-functional}
\end{cases}
\end{equation}
The onset frequency of the plateau follows the scaling law $\omega_* \propto \left(\phi-\phi_J \right)^{ \left(\alpha - 1 \right) /2 }$, while the cutoff frequency follows $ \omega_e \propto \left(\phi-\phi_J \right)^{\left(\alpha - 2\right)/2}$.
The overall factor $1/\omega_e$ ensures the normalization of $D(\omega)$. 
Note that we omitted the deviation from the non-Debye scaling law at the lowest frequency as it affects only the lowest frequency behavior of the complex modulus.
        
By substituting Eq.~(\ref{eq:vDOS-functional}) into Eq.~(\ref{eq:modulus-dos}) and performing the integral, we obtain a closed form of the complex modulus (see Methods). 
The final result is: 
\begin{equation}
G^{\prime \prime}(\omega) = 
\begin{cases}
\omega
& \left(\omega_0 \ll \omega\right) \\
\sqrt{\omega_0 \omega} 
& \left(\omega \ll \omega_0 \right), 
\end{cases}
\label{eq:scaling-ms}
\end{equation}
and
\begin{equation}
G^{\prime}(\omega) = 
\begin{cases}
\omega_0
& \left( \omega_0 \ll \omega \right) \\
\sqrt{\omega_0 \omega} 
& \left( \omega_c \ll \omega \ll \omega_0 \right) \\
\sqrt{\omega_0 \omega_c}
& \left(\omega \ll \omega_c \right),
\end{cases}
\label{eq:scaling-ms2}
\end{equation}
where the characteristic frequencies are given by $\omega_0 = \omega_e^2$ and $\omega_c = \omega_*^2$. 
The high- and intermediate-frequency parts ($\omega_c \ll \omega$) of $G^*(\omega)$ are controlled by the plateau in the vDOS, while the low-frequency part ($\omega \ll \omega_c$) is by the non-Debye scaling law of the vDOS.
These scaling functions are schematically shown in Fig.~\ref{fgr:concept}b.
These are perfectly consistent with our numerical results, except for the subtle deviation at the lowest frequency corresponding to the deviation from the non-Debye scaling law.

Several remarks are in order. 
First, our results reproduced the anomalous viscous loss $G^{\prime} \propto \omega^{0}, G^{\prime \prime} \propto \sqrt{\omega}$ at $\omega \ll \omega_c$, in contrast to the previous work~\cite{tighe2011}.
This is because the non-Debye scaling law of $D(\omega)$ is responsible for the anomalous viscous loss, and the previous work omitted it.
Second, since the non-Debye scaling law is originated from the marginal stability of the system, we can conclude that the anomalous viscous loss is also a consequence of the marginal stability. 
Third, the scaling laws of the complex modulus Eqs.~(\ref{eq:scaling-ms}) and (\ref{eq:scaling-ms2}) are independent of the exponent $\alpha$ that controls the interparticle interactions. 
This is natural because the functional form of $D(\omega)$ is independent of $\alpha$, apart from the density dependence of $\omega_*$ and $\omega_e$. 
This indicates the robustness of the anomalous viscous loss, which will be discussed later in a broader context.
Finally, the characteristic frequencies in the complex modulus are square of those  in the vDOS ($\omega_0 = \omega_e^2, \omega_c = \omega_*^2$). 
This stems from the time scale relation between the inertial dynamics in vibration and the overdamped dynamics in viscoelasticity. 
Thanks to this square, the scaling laws in the complex modulus have broader scaling regions of $\omega$, which makes their experimental and numerical observations easier. 

In summary, our theoretical analysis explains the complex modulus of the emulsions; in particular, it establishes a direct link between the anomalous viscous loss in the complex modulus and the non-Debye scaling law or, equivalently, the boson peak in the vibrations.

\section*{Scaling analysis of the experimental data}

According to Eqs.~(\ref{eq:scaling-ms}) and (\ref{eq:scaling-ms2}), the complex modulus obeys the following scaling functions: 
\begin{equation}
\frac{G^{\prime \prime}}{A \omega^{1/2}} = 1 + \left(\frac{\omega}{\omega_0}\right)^{1/2}, \ \ \ 
\frac{G^{\prime}}{B} = 1 + \left(\frac{\omega}{\omega_c}\right)^{n}.
\label{eq:fitting}
\end{equation}
The density dependence of the parameters are: $A \propto \left(\phi-\phi_J\right)^{\left(\alpha-2\right)/2}$, $B \propto \left(\phi-\phi_J\right)^{\alpha-3/2}$, $\omega_0 \propto \left(\phi-\phi_J\right)^{\alpha-2}$, $\omega_c \propto \left(\phi-\phi_J\right)^{\alpha-1}$, and $n=1/2$. 
We test this prediction by fitting the experimentally measured complex modulus, using $A, B, \omega_0, \omega_c, n$ as density-dependent fitting parameters.
Figure~\ref{fgr:scaling-modulus}ab shows the result. 
All the experimentally measured loss and storage moduli collapse well onto the scaling functions.
Figure~\ref{fgr:scaling-modulus} also shows the obtained fitting parameters, which coincide well with the theoretical predictions for $\phi > \phi_J$.
We may expect these results for $\phi > \phi_J$ based on the quantitative agreement in Fig.~\ref{fgr:exp-model}.

A remarkable feature of Fig.~\ref{fgr:scaling-modulus} is that the scaling functions work well even for $\phi < \phi_J$, where our theory is not a priori applicable.
In $\phi<\phi_J$, the thermal fluctuation is crucial, and the dynamics becomes inherently nonlinear due to the thermal collisions between particles.
The collapse suggests that the present theory can apply to this situation with minor modifications. 
One way to implement the thermal fluctuations in the soft sphere model is to introduce an effective two-body potential stemming from the thermal collisions~\cite{brito2009,Altieri2016}. 
In this approach, the strength of the effective potential becomes an order of $k_{\rm B} T$~\cite{brito2009,Altieri2016,Arceri2020}, which is much smaller than $\sigma R^2$. 
This is consistent with the fact that the parameters $A, B, \omega_0, \omega_c$ significantly decrease in $\phi < \phi_J$.

\begin{figure}[t]
\includegraphics[width=\linewidth]{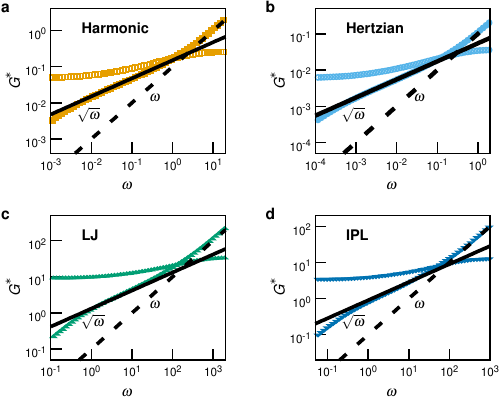}
\caption{
{\bf The anomalous viscous loss in varieties of the models.} 
{\bf a.b.c.d.} 
The numerically computed complex modulus of the packings of the harmonic spheres ({\bf a}), the Hertzian spheres ({\bf b}), the Lennard-Jones particles ({\bf c}), and the inverse-power-law particles ({\bf d}). 
The solid and dashed lines represent $G^{\prime \prime} \propto \sqrt{\omega}$ and $G^{\prime \prime} \propto \omega$. 
The anomalous viscous loss ($G^{\prime} \propto \omega^0, G^{\prime \prime} \propto \sqrt{\omega}$) emerges in all the models in a low-frequency regime.
}
\label{fig:universality}
\end{figure}

\section*{Universality of the anomalous viscous loss}

We found a link between the anomalous viscous loss and the non-Debye scaling law or the boson peak. 
The boson peak is a highly universal vibrational property of amorphous solids~\cite{phillips1981amorphous}. 
The associated non-Debye scaling law has been predicted in a general setting of the heterogeneous elasticity theory, a coarse-grained model of amorphous solids~\cite{Schirmacher2007}.
This suggests that the anomalous viscous loss is also a universal property of soft jammed solids. 
Here, we discuss this point by studying the complex modulus of varieties of model soft jammed solids. 
The details of the models are summarized in Methods. 
       
Figure~\ref{fig:universality}ab shows $G^*$ of the harmonic and Hertzian spheres, standard models for the jammed foams and granular materials.
Clearly, the anomalous viscous loss ($G^{\prime} \propto \omega^0, G^{\prime \prime} \propto \sqrt{\omega}$) emerges in a low-frequency regime. 
Note that the deviation at the lowest frequency is due to the deviation from the non-Debye scaling law at the lowest frequency, as discussed earlier.
Figure~\ref{fig:universality}cd shows $G^*$ of the Lennard-Jones and the inverse-power-law particles.
These models are standard in the theoretical studies of amorphous solids and can be viewed as the simplest models of colloidal particles with and without attractive interactions.
These models do not have a clear jamming transition, as the interactions do not have a short-range cutoff.
Again, the anomalous viscous loss emerges in a low-frequency regime. 
These results strongly suggest that the anomalous viscous loss is universal in various soft jammed solids as it comes from another universal property, the boson peak.

\section*{Summary and Discussion}

In this work, we integrated experimental and theoretical approaches to achieve a microscopic and quantitative understanding of the linear viscoelasticity of soft jammed solids. 
We now summarize and discuss our three key results. 

First, the soft sphere model quantitatively reproduced the experimental complex modulus of the concentrated oil-in-water emulsions at $\phi > \phi_J$.
This is remarkable as the model is described with a microscopic Hamiltonian with the measured (not adjustable) microscopic parameters, which is usually regarded as too simple compared to the real experimental system.
Before this study, explanations for the anomalous viscous loss were phenomenological~\cite{liu1996}.
Without a microscopic basis, such a theory neither yielded quantitative predictions nor a way to prove their validity.
The strategy in this work will play a decisive role in quantitative understanding of the viscoelasticity of other soft jammed solids, such as foams and microgels.

Second, the anomalous viscous loss $G^{\prime} \propto \omega^0, G^{\prime \prime} \propto \sqrt{\omega}$ is a direct consequence of the boson peak, a universal vibrational property of amorphous solids.
This means that the anomalous viscous loss is originated from the marginal stability of soft jammed solids. 
This link opens a new way to verify and discuss the marginal stability of real soft jammed solids by experimental measurements of the viscoelasticity~\cite{Lin2016}.

Finally, our results strongly suggest the universality of the anomalous viscous loss far beyond the realm covered by the jamming theory of soft spheres. 
Our simulations revealed the anomalous viscous loss in models with various interparticle interactions, even in those without the jamming transition. 
We attribute this universality to the universality of the boson peak in amorphous solids. 
Furthermore, our experimental $G^*$ followed the theoretical scaling functions even at $\phi < \phi_J$, where the thermal fluctuation is strong and the present theory is not a priori applicable. 
In fact, the anomalous viscous loss is observed in diverse glassy suspensions, in which thermal fluctuations play pivotal roles. 
Our recent unpublished study shows that even hard sphere suspensions exhibit similar anomalous viscous loss. 
It is thus desirable to extend the present theory to this density regime, where the dynamics becomes inherently unharmonic.
The future progress in this direction will further reveal the universality of the anomalous viscous loss in general glassy systems. 

\begin{acknowledgments}
We thank H.~Mizuno for insightful discussions, and S. Inokuchi, K. Nishi and M. Annaka for their technical supports. 
This work was supported by 
Hosokawa Powder Technology Foundation Grant Number HPTF21509,
JST SPRING Grant Number JPMJSP2108,
and JSPS KAKENHI Grant Numbers JP20H01868, JP20H00128, JP21H01048, JP22H04848, JP22K03552, JP23KJ0368.
\end{acknowledgments}

\section*{Methods}

\subsection*{Experimentl methods}

    The complex modulus $G^*(\omega)$ of concentrated emulsion was determined using passive microrheology. A laser light ($\lambda$ = 830nm, LuxX830-140, Omicron, Germany) was tightly focused onto a melamine particle (radius $a = 1\mu$m, Sigma-Aldrich) dispersed in the specimen.
    The spontaneous thermal fluctuation $u(t)$ was detected with a quadrant photodiode~\cite{schnurr1997}.
    Using the fluctuation-dissipation theorem (FDT), the complex response function $\alpha^*(\omega) = \alpha^{\prime} (\omega) - i \alpha^{\prime \prime} (\omega)$ was deduced from $u(t)$.
    $G^*$ was then obtained from the generalized Stokes' formula $G^*(\omega)=1 / 6 \pi a \alpha^*(\omega)$.

    We employed two methods to determine $\alpha^*(\omega)$.
    Firstly, in what we term the frequency-domain method, we calculated the power spectral density $\langle \abs{\tilde{u}(\omega)}^2 \rangle $ from $u(t)$.
    The FDT in the frequency domain $\alpha^{\prime \prime}(\omega) = \omega \langle \abs{\tilde{u}(\omega)}^2 \rangle/2k_{\rm B} T$ was then employed to obtain the imaginary part of the response function $\alpha^{\prime \prime} (\omega)$.
    The real part, $\alpha^{\prime}(\omega)$, was derived from $\alpha^{\prime \prime}(\omega)$ using the Kramers-Kronig relationship.
    In the second method, $\alpha^*(\omega)$ was deduced from the mean squared displacement $M(t)=\langle \abs{u(t)-u(0)}^2 \rangle$ using the FDT in time domain $\alpha(t) = (dM(t) / dt) / 2k_{\rm B} T$~\cite{nishi2018}.
    Here, $\alpha(t)$ denotes the inverse Fourier transform of $\alpha^*(\omega)$.

    Various methods exist to estimate the response function from $u(t)$.
    Despite essentially being based on the same theory, there exist technical differences in their arbitrariness and analysis uniqueness, noise introduction, and the computational cost, etc.
    The aforementioned methods we chose don't necessitate fitting to arbitrary functions or approximations.
    The frequency-domain method provides the most accurate estimation of $\alpha^{\prime \prime}(\omega)$ over the entire range of frequencies measured.
    However, Kramers-Kronig integral is truncated due to the experimental bandwidth limitations, introducing errors in $\alpha^{\prime}(\omega)$ at high frequencies. Whereas the time-domain method reduces such errors, the necessity to apply a window function for the Fourier transform of $\alpha(t)$ introduces errors at low frequencies.
    Besides, this method is computationally demanding especially when processing large data required for broad bandwidth analyses.
    Given these considerations, we decided to combine these two methods.
    In particular, $G^*(\omega)$ values above and below a certain threshold $\omega_{\rm th}$ were provided by the frequency-domain and time-domain methods, respectively.
    $\omega_{\rm th}$ was chosen as $4 \times 10^2$ rad/s for $\phi = 0.60-0.62$ and $6 \times 10^3$ rad/s at $\phi = 0.63-0.67$.

\subsection*{Experimental materials}

    The concentrated O/W emulsion was prepared with PMHS (polymethylhydrosiloxane, Mn = 1700-3200, Sigma-Aldrich) as the oil phase.
    For the water phase, an aqueous solution of the polymer surfactant Pluronic (F108, Mn = 14600, Sigma-Aldrich), DMAC (N, N-Dimethylacetamide, Sigma-Aldrich), and Formamide (Sigma-Aldrich) were mixed in a volume ratio of 6:4:1, using a Vortex genius 3 (IKA).
    The inclusion of F108, at a final concentration of 1 wt\%, stabilizes the emulsion by inducing repulsive interaction between oil droplets~\cite{senbil2019}.
    After adding PMHS to the mixture, the sample was sonicated by an ultrasonic homogenizer (microsonXL2000, MISONIX Inc., USA) on ice, using the strength levels 1 and 2 for 20 seconds each. 
    The mean radius of droplets ($R =2.4 \times 10^{-7}$ m) and the polydispersity (Pa $\equiv$ standard deviation/mean radius = 0.35) were determined by dynamic light scattering (ALV 5000/E/EPP, Langen, Germany with a He-Ne laser $\lambda$ = 632.8 nm).
    The interfacial tension $(\sigma=8.8 \times 10^{-3}$ N/m) was determined by gravimetric drop method~\cite{uematsu2018}.
    Samples were sealed in a chamber consisting of a glass slide and cover glass laminated with a polyimide tape (thickness 145 $\mu$m).
    Passive microrheology was performed in a round space created by a hole in the tape, with a diameter of 5.5 mm.

\subsection*{Soft sphere model}
    
Our central model is $N$ particles in a cubic box of volume $V$ with periodic boundary conditions, where the particles interact through the potential Eq.~(\ref{eq:potential}). 
This potential was introduced in Ref.~\cite{lacasse1996}, tuned to reproduce the osmotic pressure and the static shear modulus of concentrated emulsions.
In the original paper, the parameters $\alpha$ and $\beta$ were allowed to depend on the contact number of the particles.
However, because the contact number does not vary much in the density range we studied, we set $\alpha = 2.23$ and $\beta=0.26$, which are accurate enough near the jamming point.

We generated mechanically stable packings at the desired packing fractions by the following procedure.
First, we randomly place particles in the box at $\phi=0.50$.
Then, we slightly increase the packing fraction by decreasing the box size and minimize the potential energy by the optimization algorithm FIRE~\cite{bitzek2006FIRE}.
We iteratively continue this compression and minimization protocol until $\phi$ reaches the desired values.
We prepared 10 independent samples with $N=8000$, and took an average of observables over the samples.
We also studied $N=2000$ and 4000 cases for the highest density to verify that the finite size effect is negligible in the focused frequency regime.

\subsection*{Linear response formalism}

We summarize the method to compute the complex modulus of the soft sphere model in the microrheology setting~\cite{hara2023}. 
Since the complex modulus is a linear response property, we can linearize the equation of motion Eq.~(\ref{eq:eom}) in terms of the displacements of the particles. 
Hereafter, we denote the displacement vector of particle $i$ by $\vec{u}_i$, and use a compact notation $\ket{u}\equiv(\vec{u}_1^{\mathrm{T}}, \cdots, \vec{u}_N^{\mathrm{T}})^{\mathrm{T}}$. 

To express the left-hand side of Eq.~(\ref{eq:eom}) in a compact form, we introduce the $3N \times 3N$ dissipation matrix $\bm{C}$ whose $(i, j)$ component is given by
\begin{equation}
C_{ij} =
   \begin{cases}
        z_i C_0 \bm{I} & (i=j) \\
        - C_0 \bm{I} & (\text{$ij$ are in contact}) \\
       \bm{O} & (\text{otherwise}),
  \end{cases}
\end{equation}
where $z_i$ is the contact number of particle $i$, and $\bm{I}, \bm{O}$ are the $3 \times 3$ unit and zero matrices, respectively.
For the right-hand side of Eq.~(\ref{eq:eom}), we expand the potential energy as $U = U_0 + \frac{1}{2} \expval{\bm{\mathcal{M}}}{u}$, where 
\begin{equation}
\bm{\mathcal{M}} = \pdv{U}{\vec{u}_j}{\vec{u}_i} 
 \label{eq:hessian}
\end{equation}
is the $3N \times 3N$ Hessian matrix. 
Then, the linearized equation of motion can be written as 
\begin{equation}
   \bm{C} \ket{\dot{u}} = - \bm{\mathcal{M}} \ket{u} + \ket{F}. 
\end{equation}
We apply the external force $F_{p, \gamma}$ to particle $p$ in the direction $\gamma$, hence $\ket{F} = (\vec{0}^{\mathrm{T}}, \cdots, F_{p, \gamma} , \cdots, \vec{0}^{\mathrm{T}})^{\mathrm{T}}$. 

We can solve this equation by the Fourier transformation: 
\begin{equation}
     \ket{\hat{u}(\omega)} = \bm{G}(\omega) \ket{\hat{F}(\omega)}, 
\end{equation}
where $\bm{G}(\omega) \equiv (\mathcal{M} + i \omega \bm{C})^{-1}$ is the Green's function. 
The response function is a diagonal element of the Green's function $G_{p\gamma, p\gamma}(\omega)$ as it expresses the response of the probe particle. 
We need to invert $\mathcal{M} + i \omega \bm{C}$ to obtain a closed form of $\bm{G}(\omega)$.
We can achieve this by introducing the auxiliary matrix $\bm{\tilde{\mathcal{M}}}=\bm{C}^{1/2}\bm{\mathcal{M}}\bm{C}^{1/2}$.
After some algebra~\cite{hara2023}, we obtain 
\begin{align}
\alpha^*(\omega) 
& = \sum_{m n \mu \nu k} (\bm{C^{-\frac{1} {2}}})_{p\gamma,m\mu}  \frac{\tilde{e}_{m, \mu}(\omega_k) \tilde{e}_{n, \nu}(\omega_k) }{\omega_k^2 + i \omega} (\bm{C^{-\frac{1}{2}}})_{n\nu, p\gamma},
 \label{eq:response-function-LR}
\end{align}
where $\omega_k$ and $\ket{\tilde{e}(\omega_k)}$ are the $k$-th eigenfrequency and eigenvector of $\bm{\tilde{\mathcal{M}}}$. 
Note that we may stick to the eigenvalues $\lambda_k$ here, instead of using the eigenfrequencies $\omega_k = \sqrt{\lambda_k}$. 
Using the eigenfrequencies corresponds to introducing the hypothetical inertial motion of the particles mentioned in the main text.   

In practice, we numerically diagonalized $\bm{\tilde{\mathcal{M}}}$ for the given mechanically stable packings, calculated the response function $\alpha^*(\omega)$ by Eq.~(\ref{eq:response-function-LR}), and converted it to the complex modulus via the generalized Stokes' formula $G^*(\omega)=1 / 6 \pi R \alpha^*(\omega)$.

In this approach, we assumed that the probe particle and the other particles are identical and regarded a randomly chosen particle as the probe.
Our previous work~\cite{hara2023} proved the validity of this assumption: the computed complex modulus quantitively agrees with the macroscopic complex modulus, and moreover, it does not change much even if we increase the size of the probe particle. 
This assumption enabled us to take an average of $G^*$ over the choice of the probe particle, namely the indices $p$ and $\gamma$, to improve the statistics.

\subsection*{Details of the theoretical analysis}

We derive Eqs.~(\ref{eq:modulus-dos}), (\ref{eq:scaling-ms}), and (\ref{eq:scaling-ms2}) in the section of theoretical analysis. 
Since we focus on the scaling behaviors, we omit unimportant $O(1)$ quantities in this section. 

\begin{figure}[h]
\includegraphics[width=\linewidth]{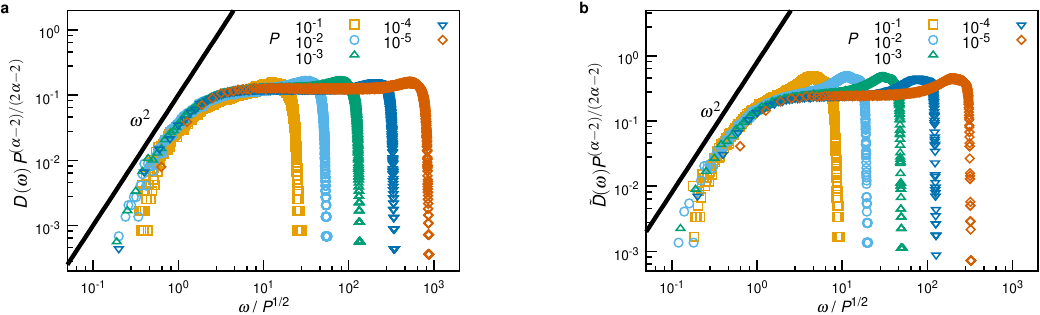}
\caption{
{\bf Scaling analysis of the vibrational density of states.} 
{\bf a.} $P^{(\alpha-2)/2(\alpha-1)} D(\omega)$ is plotted against $\omega /P^{1/2}$ for the soft sphere model, where $P$ is the pressure of the packings. 
The results for $P=10^{-5} - 10^{-1}$ with $N=8000$ are included.
According to Eq.~(\ref{eq:vDOS-functional}), the vDOS of the soft sphere model obeys the scaling relation $\omega_e D(\omega) = f(\omega/\omega_*)$ with $f(x) \propto x^2$ at $x < 1$.
Because the pressure follows the scaling relation $P \propto (\phi - \phi_J)^{\alpha -1}$, we can express the characteristic frequencies as $\omega_e \propto P^{(\alpha-2)/(2\alpha-2)}$ and $\omega_* \propto P^{1/2}$. 
Therefore, the scaling relation can be rewritten as $P^{(\alpha-2)/(2\alpha-2)} D(\omega) = g(\omega/P^{1/2})$.
The data collapse in the figure confirms this scaling relation of $D(\omega)$.
{\bf b.} Same as {\bf a} but for $\tilde{D}(\omega)$, the density of eigenfrequencies of $\bm{\tilde{\mathcal{M}}}$. 
The data collapse confirms that $D(\omega)$ and $\tilde{D}(\omega)$ share the same scaling relation.
}
\label{fgr:scaling-dos}
\end{figure}

Our starting point is Eq.~(\ref{eq:response-function-LR}). 
For simplicity, we take the average of it over $p$ and $\gamma$, the choice of the probe particle. 
Then, we can simplify this formula into Eq.~(\ref{eq:modulus-dos}) as follows.
First, because the eigenvectors in the relevant frequency regime are spatially disordered and extended, the elements of the eigenvectors are virtually random and $\tilde{e}_{m, \mu}(\omega_k) \tilde{e}_{n, \nu}(\omega_k)$ can be approximated as $\frac{1}{3N} \delta_{m, n} \delta_{\mu, \nu}$. 
Here, the factor $\frac{1}{3N}$ ensures the normalization of the eigenvectors. 
Second, the eigenvalues of $\bm{C}$ do not have any critical property near the jamming transition.
As a result, we can expect $\frac{1}{3N} \sum_{p\gamma} (\bm{C^{-1}})_{p\gamma,p\gamma} = O(1)$ for any density, which we have confirmed numerically. 
Therefore, Eq.~(\ref{eq:response-function-LR}) can be recast into 
\begin{equation}
\alpha^*(\omega) = \int d\omega^{\prime} \frac{\tilde{D}(\omega^{\prime})}{\left(\omega^{\prime}\right)^2 + i \omega}, 
\label{eq:modulus-dos0}
\end{equation}
where $\tilde{D}(\omega)$ is the density of eigenfrequencies of $\bm{\tilde{\mathcal{M}}}$. 
In principle, $\tilde{D}(\omega)$ is different from the vDOS $D(\omega)$, which is the density of eigenfrequencies of $\bm{\mathcal{M}}$. 
However, we numerically checked that $\tilde{D}(\omega)$ and $D(\omega)$ share the same scaling relations in the wide range of densities (Extended data Fig.~\ref{fgr:scaling-dos}).
Therefore, we can safely replace $\tilde{D}(\omega^{\prime})$ with $D(\omega^{\prime})$ to obtain Eq.~(\ref{eq:modulus-dos}). 

Eqs.~(\ref{eq:scaling-ms}) and (\ref{eq:scaling-ms2}) can be derived as follows. 
Substituting Eq.~(\ref{eq:vDOS-functional}) into Eq.~(\ref{eq:modulus-dos}) and introducing the variable $x = \omega'/\sqrt{\omega}$, we obtain 
\begin{equation}
\frac{\sqrt{\omega_0 \omega}}{G^*} = 
\int^{\sqrt{\frac{\omega_0}{\omega}}}_{\sqrt{\frac{\omega_c}{\omega}}} \frac{dx}{x^2 + i} 
+ \frac{\omega}{\omega_c} \int^{\sqrt{\frac{\omega_c}{\omega}}}_0 \frac{x^2 dx}{x^2 + i}, 
\label{eq:integrals}
\end{equation}
where we introduced two natural frequency scales $\omega_c = \omega_*^2$ and $\omega_0 = \omega_e^2$. 
The first integral is the contribution from the plateau, and the second is the non-Debye scaling in the vDOS.
Now, we can efficiently perform an asymptotic evaluation of the integrals using $(x^2 +i)^{-1} \to x^2 -i$ at $x \ll 1$ and $\to x^{-2} -i x^{-4}$ at $x \gg 1$. \\
\noindent
(i) High-frequency regime ($\omega_c \ll \omega_0 \ll \omega$): 
The first term asymptotically gives $(\omega_0/\omega)^{3/2} - i(\omega_0/\omega)^{1/2}$. 
The second term gives $(\omega_c/\omega)^{3/2} - i(\omega_c/\omega)^{1/2}$, and hence the first is dominant. 
Inverting the first term, we obtain $G^* = \omega_0 + i \omega$. \\
\noindent
(ii) Intermediate-frequency regime ($\omega_c \ll \omega \ll \omega_0$): 
The first term asymptotically gives $1 -i$ while the second term gives $(\omega_c/\omega)^{3/2} - i(\omega_c/\omega)^{1/2}$. 
Again, the first is dominant, and we obtain $G^* = \sqrt{\omega_0 \omega} + i \sqrt{\omega_0 \omega}$. \\
\noindent
(iii) Low-frequency regime ($\omega \ll \omega_c \ll \omega_0$): 
The first term asymptotically gives $(\omega/\omega_c)^{1/2} - i(\omega/\omega_c)^{3/2}$ while the second term gives $(\omega/\omega_c)^{1/2} - i(\omega/\omega_c)$. 
The first and second terms are in the same order for the real part, whereas the second is dominant for the imaginary part.
Therefore, inverting $(\omega/\omega_c)^{1/2} - i(\omega/\omega_c)$, we obtain $G^* = \sqrt{\omega_0 \omega_c} + i \sqrt{\omega_0 \omega}$. \\
\noindent
Collecting all three cases, we obtain 
\begin{equation}
  G^*(\omega) = 
  \begin{cases}
      \omega_0 + i \omega
      & \left( \omega_c \ll \omega_0 \ll \omega \right) \\
      \sqrt{\omega_0 \omega} + i \sqrt{\omega_0 \omega}
      & \left( \omega_c \ll \omega \ll \omega_0 \right) \\
      \sqrt{\omega_0 \omega_c} + i \sqrt{\omega_0 \omega}
      & \left(\omega \ll \omega_c\right). 
  \end{cases}
\label{eq:scaling-G}
\end{equation}
Interestingly, the imaginary parts in the intermediate- and low-frequency regimes are in the same order of $\sqrt{\omega_0 \omega}$. 
Hence, we can merge them to obtain Eqs.~(\ref{eq:scaling-ms}) and (\ref{eq:scaling-ms2}).

\subsection*{Additional models}

We additionally studied four models with different interparticle interactions.
The interaction potentials for the harmonic spheres and the Hertzian spheres are 
\begin{equation}
  v(r) = \frac{\epsilon}{\gamma} \left(1 - \frac{r}{D}\right)^{\gamma}\Theta\left(1 - \frac{r}{D}\right),
\end{equation}
where $\gamma=2$ and $\gamma=5/2$ correspond to the harmonic and the Hertzian, respectively.
Using the method described above, we generated mechanically stable packings of these models at $\phi=0.67$.

The interaction potential for the Lennard-Jones particles is 
\begin{equation}
   v(r) = 4 \epsilon \left(\left(\frac{D}{r}\right)^{12} - \left(\frac{D}{r}\right)^6\right),
\end{equation}
and that for the inverse-power-law particles is 
\begin{equation}
   v(r) = \epsilon \left(\frac{D}{r}\right)^{12}. 
\end{equation}
We truncated these potentials at $r=2.5D$ and added the linear terms to ensure the continuity of the first derivative at $r=2.5D$~\cite{shimada2018}.
By the following procedure, we generated the mechanically stable packings of these models with the number density $ND^3/V=1$.
We first performed molecular dynamics simulations of these systems at a high enough temperature to obtain an equilibrium liquid state.
We then apply the FIRE algorithm to the equilibrium liquid configurations to obtain the mechanically stable packings.
As in the case of the soft sphere model, we prepared 10 independent samples with $N=8000$, and took an average of observables over the samples.

For these models, we utilized the equation of motion
\begin{equation}
   C_{0} \dv{\vec{r}_i}{t} = - \pdv{U}{\vec{r}_i} + \vec{F}_p \delta_{ip}. 
\end{equation}
Here, we adopted the Stokes dissipation for simplicity since this modification did not change the scaling laws of the complex modulus.
For Fig.~\ref{fig:universality}, the length, energy, and time units are chosen as $D$, $\epsilon$, and $C_0 D^2/\epsilon$, respectively.

\bibliography{marginal.bib}

\pagebreak
\widetext
\newpage
\begin{center}
\textbf{\large Extended data}
\end{center}

\end{document}